%% file: Template.tex
\documentclass{article}
\usepackage{spconf,amsmath,graphicx,hyperref}
\usepackage{subcaption}
\usepackage{cite}
\usepackage{booktabs}
\usepackage{multirow}

\title{When Silence Matters: The Impact of Irrelevant Audio on Text Reasoning in Large Audio-Language Models}
\name{Chen-An Li$^1$, Tzu-Han Lin$^1$, Hung-yi Lee$^{1,2}$}
\address{$^1$National Taiwan University \\
$^2$NTU Artificial Intelligence Center of Research Excellence (NTU AI-CoRE)}
\begin{document}
\ninept
\maketitle
\begin{abstract}
Large audio-language models (LALMs) unify speech and text processing, but their robustness in noisy real-world settings remains underexplored. We investigate how irrelevant audio, such as silence, synthetic noise, and environmental sounds, affects text reasoning tasks where audio is unnecessary. Across three text-based benchmarks, we find that even non-informative audio reduces accuracy and increases prediction volatility; the severity of interference scales with longer durations, higher amplitudes, and elevated decoding temperatures. Silence, often assumed neutral, destabilizes outputs as strongly as synthetic noise. While larger models show greater resilience, vulnerabilities persist across all evaluated systems. We further test mitigation strategies and find that prompting shows limited effectiveness, whereas self-consistency improves stability at the cost of increased computation. Our results reveal cross-modal interference as a key robustness challenge and highlight the need for efficient fusion strategies that preserve reasoning performance in the presence of irrelevant inputs. The code and data are publicly available at \url{https://github.com/lca0503/AudioInterference}.
\end{abstract}
\section{Introduction}
\label{sec:intro}
Large audio-language models (LALMs)~\cite{gpt4o, comanici2025gemini, lu2025desta2, liu2025voxtral, abouelenin2025phi, xu2025qwen2, goel2025audio} have shown strong performance across a variety of multimodal tasks, showing the ability to process speech and text in a unified framework~\cite{huang2025dynamicsuperb, yang25g_interspeech, sakshi2025mmau, ma2025mmar, yang-etal-2024-air, wang-etal-2025-audiobench, lin2025preliminary, yang2025towards, arora2025landscape}. However, most evaluations assume clean, modality-aligned inputs. In practice, text reasoning often requires no audio, yet deployed systems still receive streams containing silence, background noise, or incidental sounds. Intuitively, we might expect such irrelevant audio to have little or no effect on the model’s text-based reasoning. Surprisingly, our study reveals that even these non-informative signals can interfere with the fusion process and degrade performance.

Prior work has highlighted vulnerabilities of LALMs under more adversarial conditions\cite{carlini2018audio, peri-etal-2024-speechguard}. Some studies investigate injection attacks~\cite{hou2025evaluating}, while others explore cross-modal conflicts where audio and text contradict each other~\cite{wang2025audio}. Several studies have also examined how distractions affect large language models~\cite{shi2023large, chiang-lee-2024-reasoning, wang2025breaking} and vision-language models~\cite{liu2025robustness, deng2025words, cai2025diagnosing}, showing that models often struggle when exposed to irrelevant or misleading content. However, little attention has been given to the simpler but pervasive case of irrelevant audio that does not conflict with text yet degrades performance.

In this work, we ask: \textit{To what extent are LALMs vulnerable to irrelevant audio during text reasoning tasks?} To address this question, we systematically evaluate the effect of non-informative audio on established text-based benchmarks~\cite{cobbe2021training, clark2018think, hendrycks2021measuring}. In our setup, the textual input remains fixed while the audio channel varies with non-semantic perturbations such as silence, noise, or environmental sound~\cite{fonseca2021fsd50k}. We also conduct ablations, varying decoding temperature, audio duration, and noise amplitude, to study how interference strength scales across conditions.

We find that irrelevant audio lowers accuracy and alters model outputs, even when text alone is sufficient to solve the task. Surprisingly, even silence can interfere. Degradation intensifies with longer or louder noise, or with extended silence, and becomes especially pronounced at higher temperatures. Naive mitigation, such as prepending instructional phrases, proves ineffective. A simple self-consistency~\cite{wang2023selfconsistency} approach offers partial mitigation but increases test-time cost~\cite{snell2024scaling}, suggesting future work should seek more efficient solutions.  These findings establish cross-modal interference as an essential evaluation axis and highlight the need for fusion strategies that preserve reasoning against irrelevant multimodal inputs.

\begin{figure}[hbtp]
    \centering
     \vspace{-5pt}
    \includegraphics[width=0.9\linewidth]{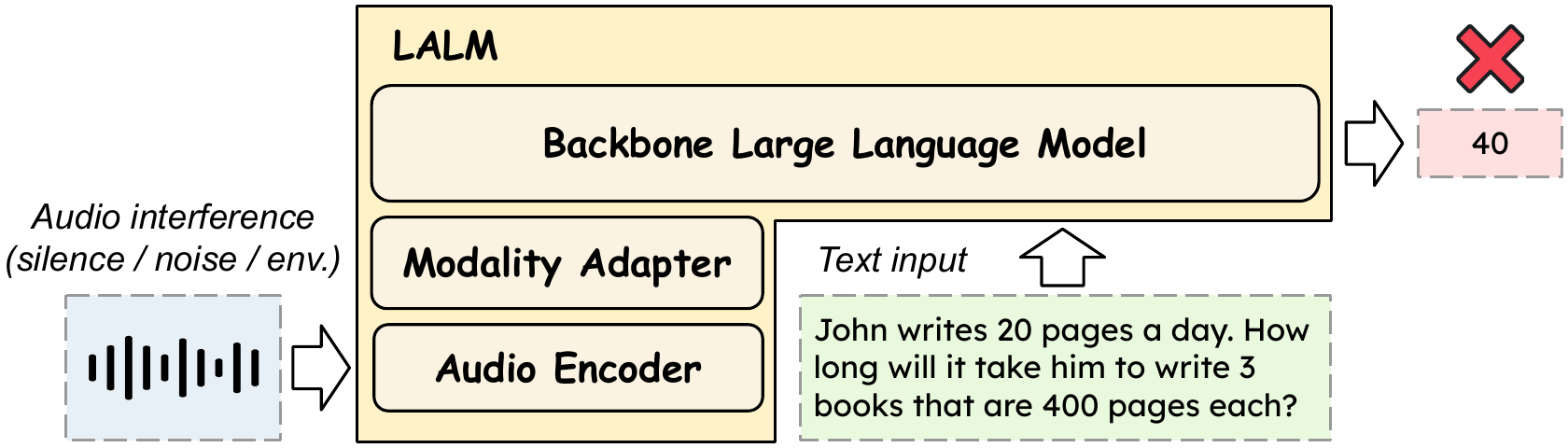}
    \vspace{-5pt}
    \caption{Illustration of cross-modal interference: irrelevant audio disrupts text-only reasoning in LALMs}
    \vspace{-15pt}
    \label{fig:framework}
\end{figure}

\section{Investigating Cross-Modal Interference}
\label{sec:investigate}

\subsection{Problem Formulation}
\label{subsec:formulation}
We analyze how large audio-language models (LALMs) handle tasks that rely only on text when the audio channel introduces irrelevant or distracting content. Figure~\ref{fig:framework} illustrates our problem setup, a text-only reasoning task with irrelevant audio signals such as silence, synthetic noise, or environmental sounds. Formally, the model generates predictions according to $\hat{y} = f_\theta(x_\text{audio}, x_\text{text}),$ where $f_\theta$ denotes the modeling process, $x_\text{audio}$ is the audio input, $x_\text{text}$ is the text input, and $\hat{y}$ is the model output.

In the normal case, the model input is $(\emptyset, x_\text{text}),$ where $\emptyset$ denotes the absence of audio. We introduce audio signals that should not affect task performance to study interference. These signals may include stretches of silence, bursts of synthetic noise, or unrelated real-world sounds such as rainfall, flowing water, or animal calls. Under interference, the model input becomes $(\delta_\text{audio}, x_\text{text}),$ where $\delta_\text{audio}$ represents non-informative audio.

By fixing $x_\text{text}$ and systematically varying $\delta_\text{audio}$, we characterize how irrelevant audio influences model behavior, measuring accuracy degradation and shifts in generated outputs.
\begin{figure*}[htbp]
    \centering
    \begin{subfigure}{0.94\linewidth}
        \centering
        \includegraphics[width=\linewidth]{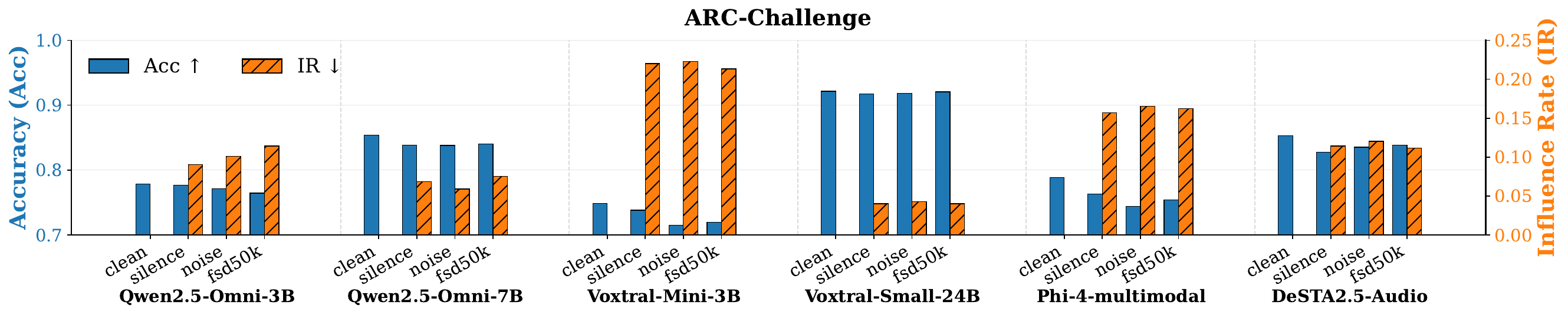}
    \end{subfigure}
    \hfill 
    \begin{subfigure}{0.94\linewidth}
        \centering
        \includegraphics[width=\linewidth]{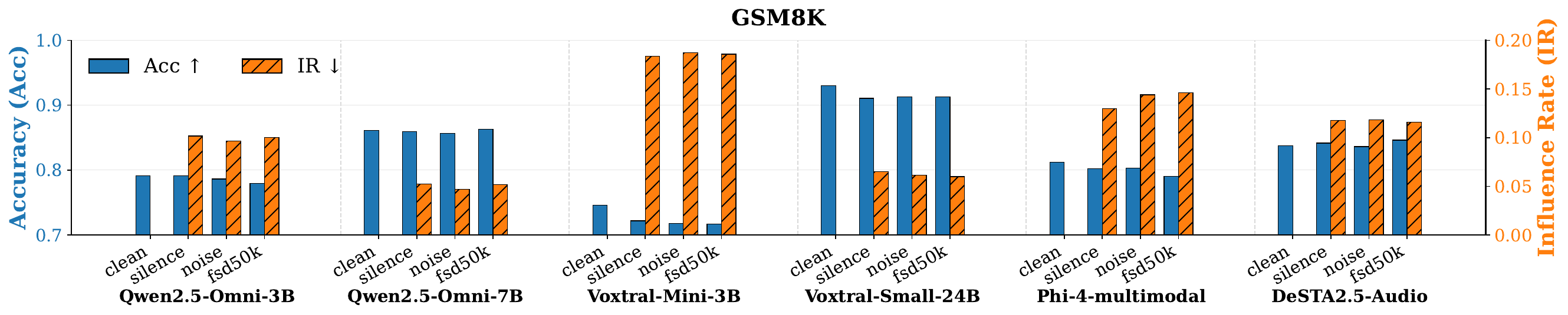}
    \end{subfigure}
    \hfill 
    \begin{subfigure}{0.94\linewidth}
        \centering
        \includegraphics[width=\linewidth]{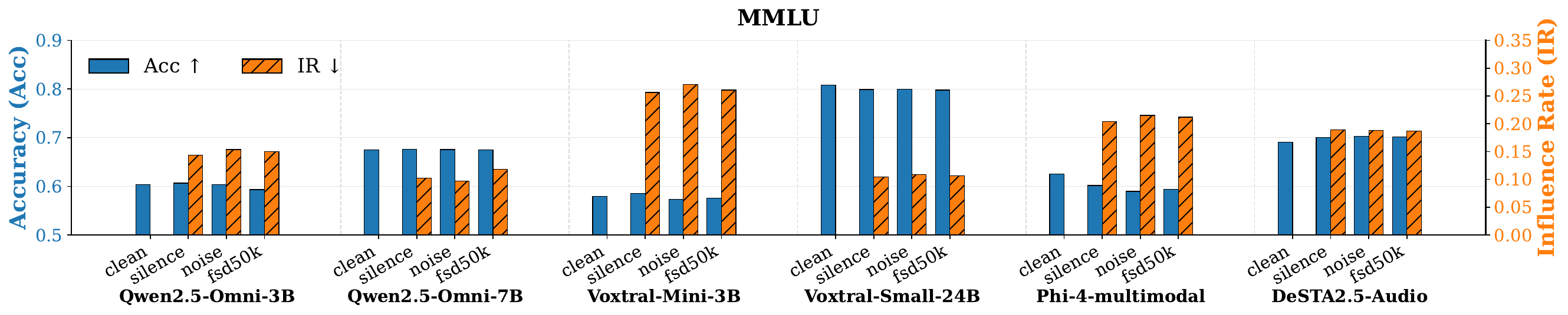}
    \end{subfigure}
    \vspace{-5pt}
    \caption{Accuracy (Acc) and Influence Rate (IR) of LALMs under cross-modal interference across benchmarks.}
    \label{fig:main}
    \vspace{-10pt}
\end{figure*}

\subsection{Experimental Setup}
\label{subsec:setup}

\noindent\textbf{Benchmarks}\quad
For tasks that rely entirely on textual reasoning, we evaluate models on three widely used benchmarks: GSM8K~\cite{cobbe2021training} for arithmetic reasoning, ARC-Challenge~\cite{clark2018think} for science question answering, and MMLU~\cite{hendrycks2021measuring} for massive multi-task language understanding. To probe robustness, we introduce three types of audio interference:  
(i) five seconds of silence,  
(ii) five seconds of synthetic Gaussian noise at \(-40\) dBFS,  
(iii) real-world audio samples drawn from FSD50K~\cite{fonseca2021fsd50k}, a diverse dataset of music, environmental sounds, and sound effects. This setup allows us to study how irrelevant audio influences text-based reasoning tasks systematically.

\noindent\textbf{Models}\quad
We evaluate a diverse set of state-of-the-art open-source LALMs to assess robustness against irrelevant audio in text reasoning tasks. Our model set includes Qwen2.5-Omni-3B, Qwen2.5-Omni-7B~\cite{xu2025qwen2}, Phi-4-Multimodal-Instruct~\cite{abouelenin2025phi}, Voxtral-Mini-3B, Voxtral-Small-24B~\cite{liu2025voxtral}, and DeSTA2.5-Audio~\cite{lu2025desta2}. These models differ in parameter size and architectural design, providing a representative view of current multimodal approaches. For most models, we adopt greedy decoding to isolate performance without the added variance of stochastic sampling, which makes the evaluation more stable under controlled perturbations. The only exception is the Voxtral series, where we follow the authors’ recommended configuration and apply nucleus sampling with temperature set to 0.2 and top-p set to 0.95. We use vLLM~\cite{kwon2023efficient} for inference, except DeSTA2.5-Audio, which uses the Transformers~\cite{wolf-etal-2020-transformers} package.

\noindent\textbf{Metrics}\quad
We adopt two evaluation metrics to assess robustness under irrelevant audio.  First, we report accuracy, defined as the proportion of correctly answered samples relative to the total number of samples in the dataset:  $\textit{Accuracy} = n_c/N$,
where $n_c$ denotes the number of correct predictions and $N$ the total number of samples, accuracy reflects overall task performance under different interference conditions.  

Second, we use the influence rate, a notion explored in other multimodal robustness studies~\cite{wang2025audio}, to quantify how often irrelevant audio changes model predictions. Let $n_{ic}$ be the number of cases where a prediction flips from incorrect to correct, and $n_{ci}$ the number of cases where it flips from correct to incorrect. We compute the influence rate as $\textit{Influence Rate} = (n_{ic} + n_{ci})/N.$
A higher value indicates greater sensitivity to irrelevant audio, regardless of whether the change improves or harms accuracy.

\subsection{Main Observation}
\label{subsec:main}
Figure~\ref{fig:main} shows that all evaluated LALMs are vulnerable to cross-modal interference. In the plots, \textit{clean} denotes clean inputs without interference, \textit{silence} adds silent audio, \textit{noise} corresponds to Gaussian noise, and \textit{fsd50k} represents a real-world audio sample from FSD50K. Across GSM8K, ARC-Challenge, and MMLU, introducing irrelevant audio consistently reduces accuracy compared to the clean setting. The absolute drops remain modest, yet they appear across all models and benchmarks, showing that even non-informative audio slightly harms performance.

The influence rate provides a clearer view of instability. Under interference, predictions frequently change, leading to noticeably higher influence rates across all conditions. Surprisingly, silence, often assumed to be a ``neutral input'', destabilizes outputs when persistent. Silence and Gaussian noise produce similar effects, while FSD50K sometimes increases the disruption, but not uniformly across tasks. This variation suggests that models do not react to one type of distractor consistently; instead, any non-informative audio has the potential to shift predictions. Accuracy alone fails to capture the full extent of instability, as outputs may change substantially even when task performance appears steady.

In addition, we observe a scaling effect when comparing models with the same architecture but different parameter sizes; larger models generally achieve better performance and exhibit reduced sensitivity to interference. In other words, larger LALMs are more robust against irrelevant audio, showing minor accuracy drops and lower influence rates under the same perturbations. Beyond scaling, we also note that some models, such as Qwen2.5-Omni, display comparatively lower volatility than others of similar size. This suggests that factors beyond parameter count, such as training data and optimization design, may influence robustness, highlighting that resilience is shaped by multiple dimensions rather than scaling alone.

Another trend visible in Figure~\ref{fig:main} is that the degree of interference differs across tasks. MMLU, which requires broad multi-domain reasoning, suffers from larger performance degradation and higher instability than more structured tasks like GSM8K arithmetic or ARC-Challenge science questions. 

Irrelevant audio lowers accuracy, raises influence rates, and introduces uneven task- and model-dependent vulnerabilities. The persistence of this effect, even under greedy decoding, emphasizes the need for more robust multimodal fusion strategies that preserve both accuracy and stability in the presence of irrelevant inputs.

\label{subsec:analysis}
\begin{figure}[htbp]
    \centering
    \begin{subfigure}{0.49\linewidth}
        \centering
        \includegraphics[width=\linewidth]{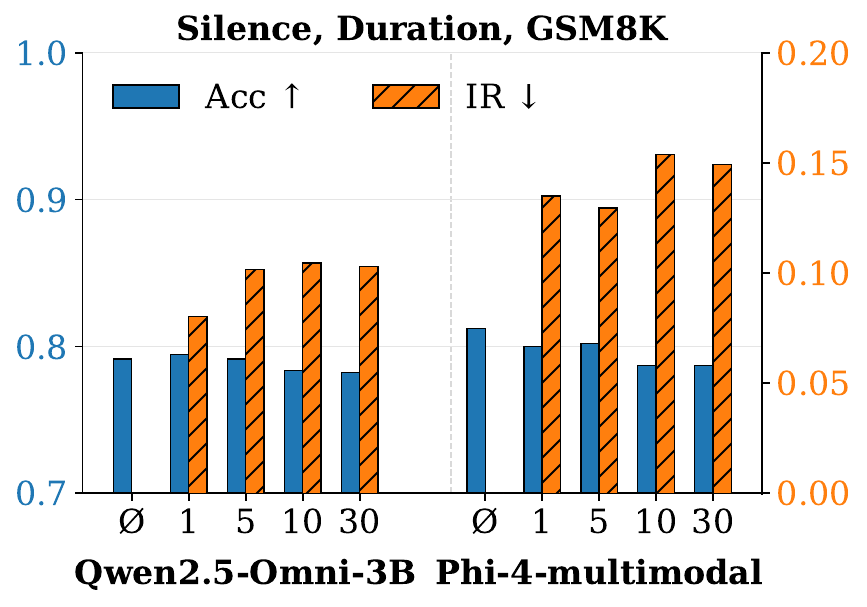}
    \end{subfigure}
    \begin{subfigure}{0.49\linewidth}
        \centering
        \includegraphics[width=\linewidth]{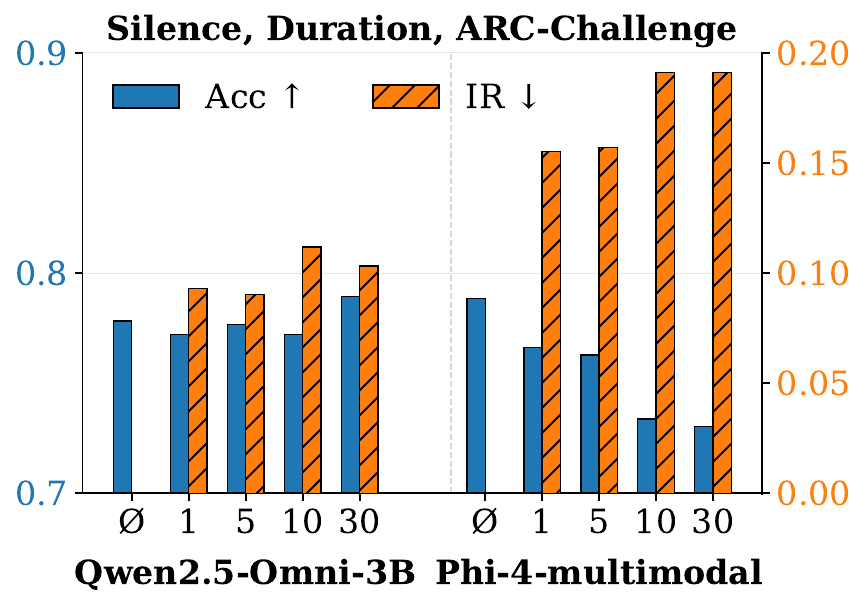}
    \end{subfigure}
    \par\vspace{5pt}
     \begin{subfigure}{0.49\linewidth}
        \centering
        \includegraphics[width=\linewidth]{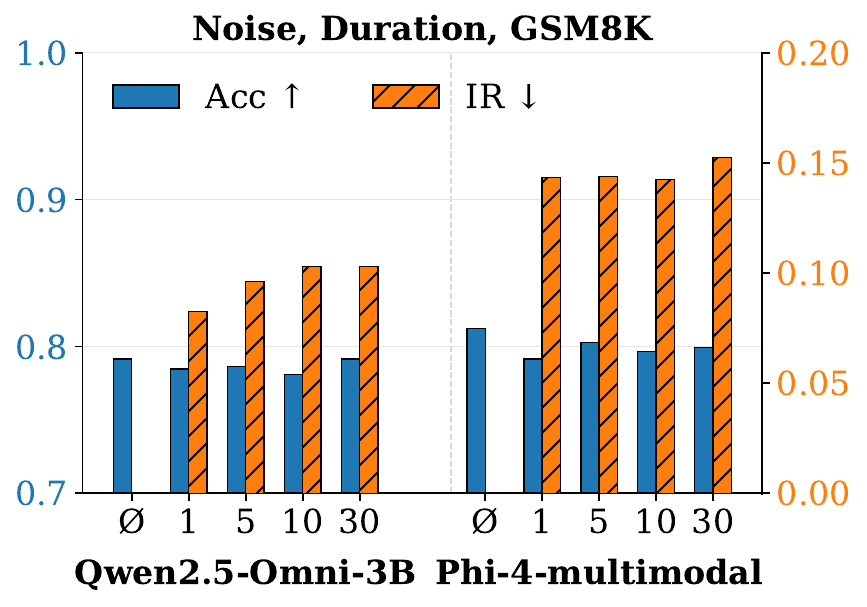}
    \end{subfigure}
    \begin{subfigure}{0.49\linewidth}
        \centering
        \includegraphics[width=\linewidth]{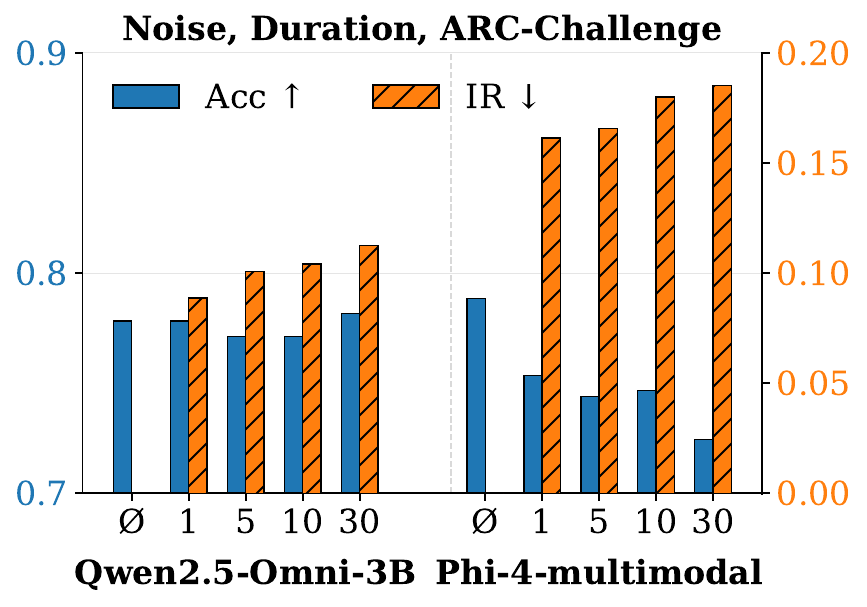}
    \end{subfigure}
    \par\vspace{5pt}
    \begin{subfigure}{0.49\linewidth}
        \centering
        \includegraphics[width=\linewidth]{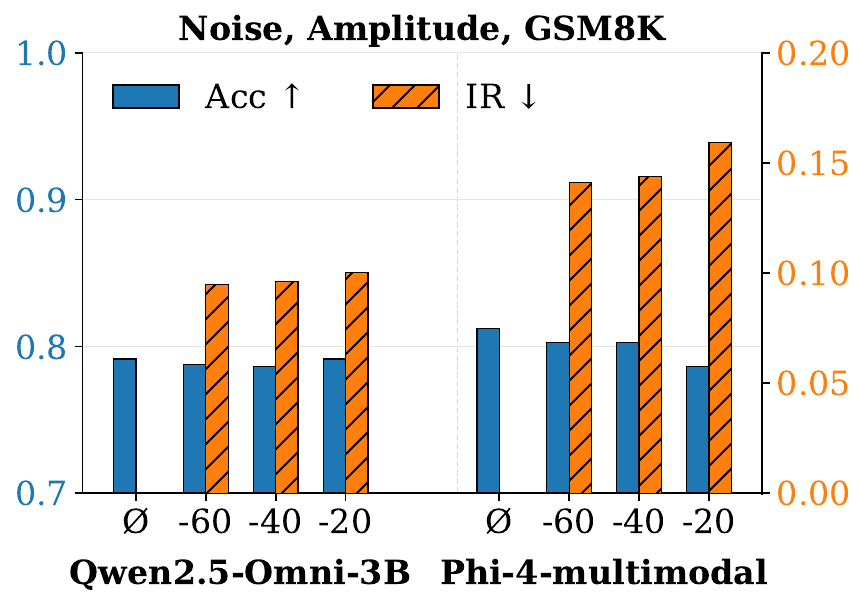}
    \end{subfigure}
    \begin{subfigure}{0.49\linewidth}
        \centering
        \includegraphics[width=\linewidth]{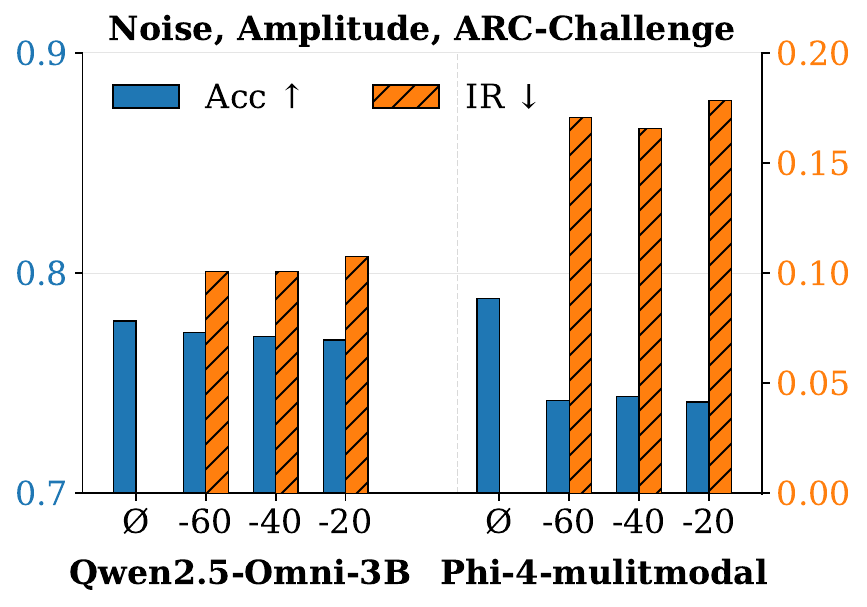}
    \end{subfigure}
    \caption{Impact of varying silence and noise durations (1, 5, 10, 30 sec) and noise amplitudes (-60, -40, -20 dBFS) on GSM8K and ARC-Challenge.}
    \label{fig:effect}
    \vspace{-10pt}
\end{figure}
\section{Analysis}
\label{sec:deeper}
\subsection{Scaling Interference Effects}
\noindent\textbf{Duration of Audio}\quad
Figure~\ref{fig:effect} illustrates the model's performance and influence rate across different durations of irrelevant audio. The x-axis indicates the duration of added audio, $\emptyset$ represents the clean baseline without any interference, while the values 1, 5, 10, and 30 denote durations of silence and Gaussian noise in seconds, respectively. As duration increases, accuracy consistently drops and influence rate rises, revealing that longer non-informative segments amplify cross-modal interference. Even silence, when extended, destabilizes reasoning, suggesting that persistent irrelevant signals are not ignored but gradually entangle in the fusion process. This scaling highlights that temporal persistence of irrelevant signals is a critical factor in LALMs' robustness.

\noindent\textbf{Amplitude of Noise}\quad
Figure~\ref{fig:effect} also demonstrates the effect of noise amplitude on model robustness. We injected 5-second Gaussian noise segments at three intensity levels, -60 dBFS, -40 dBFS, and -20 dBFS. The results indicate a clear trend in which model accuracy consistently declines as the amplitude increases and the influence rate rises. This pattern indicates that louder noise exacerbates cross-modal interference, making the model less stable and more prone to prediction shifts. Even when textual reasoning should dominate, high-intensity noise disrupts the fusion process, amplifying instability and eroding performance on benchmarks. In other words, stronger noise levels act as a more forceful distractor, directly undermining the reliability of LALMs in text reasoning tasks.

\input{Tables/cross}
\subsection{Comparative Analysis of Interference Types}
\label{subsec:cross-comparison}
Table~\ref{tab:cross} compares how model predictions change under different types of irrelevant audio by reporting the correctness change ratio. Each value reflects the proportion of samples where the model’s prediction flipped from correct to incorrect, or vice versa, when moving between two interference conditions. For instance, a value of 0.057 for ``silence/noise'' in Qwen2.5-Omni-3B means that 5.7\% of predictions changed correctness when the exact input text was paired with silence instead of Gaussian noise. For both Qwen2.5-Omni-3B and Phi-4-Multimodal, the silence–noise ratio is consistently lowest, indicating that these models treat silence and Gaussian noise as nearly equivalent. Additionally, we observe that the silence versus FSD50K comparison yields the highest ratio, with noise versus FSD50K falling in between, suggesting that FSD50K audio is perceptually closer to noise than to silence for these models.

\begin{figure}[htbp]
    \centering
    \begin{subfigure}{0.45\linewidth}
        \centering
        \includegraphics[width=\linewidth]{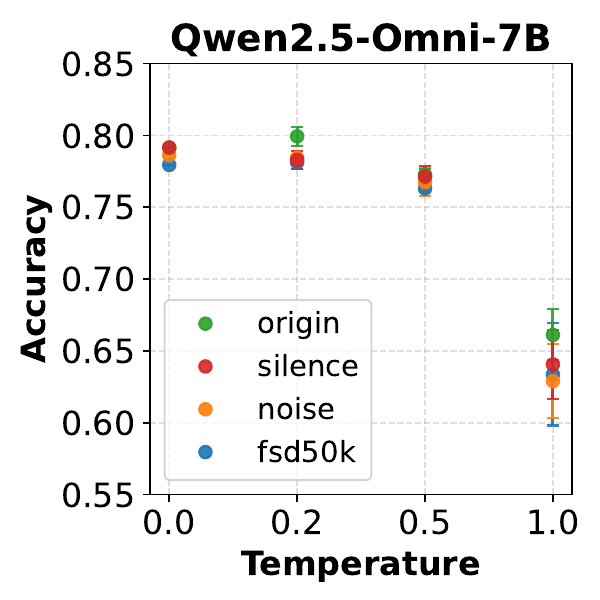}
    \end{subfigure}
    \begin{subfigure}{0.45\linewidth}
        \centering
        \includegraphics[width=\linewidth]{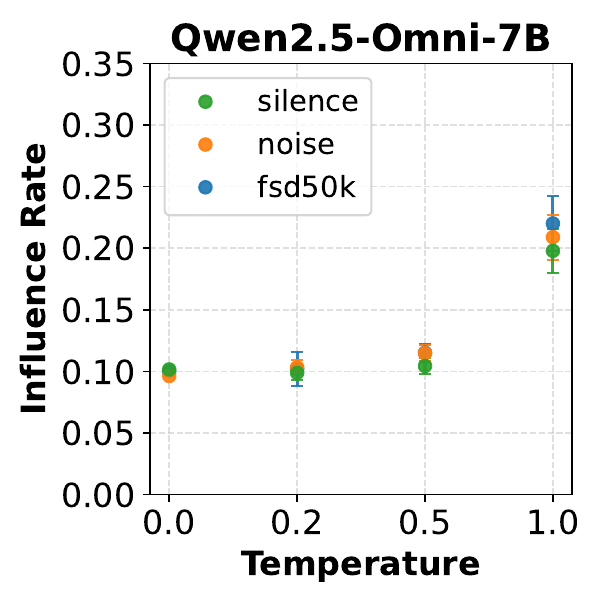}
    \end{subfigure}
    \begin{subfigure}{0.45\linewidth}
        \centering
        \includegraphics[width=\linewidth]{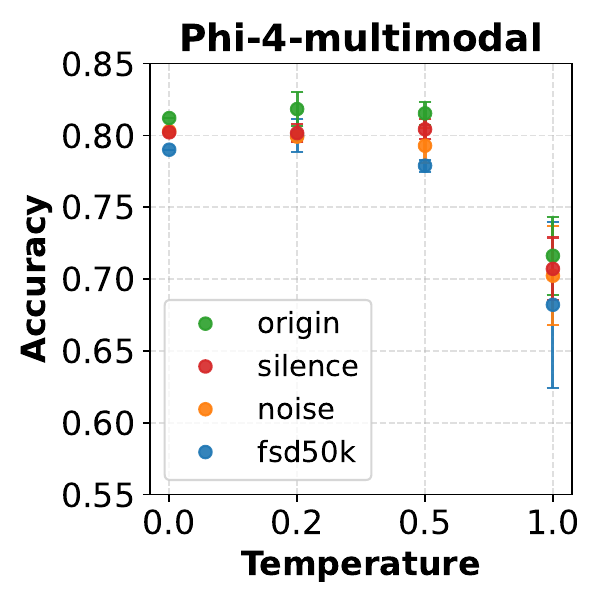}
    \end{subfigure}
    \begin{subfigure}{0.45\linewidth}
        \centering
        \includegraphics[width=\linewidth]{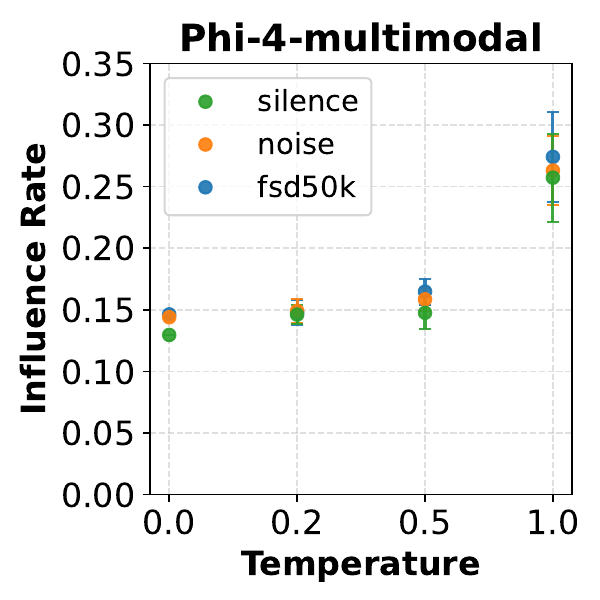}
    \end{subfigure}
    \caption{Effect of decoding temperature on accuracy and influence rate under audio interference on GSM8K. Non-greedy results are averaged over 3 seeds with standard deviation.}
    \label{fig:temp}
     \vspace{-10pt}
\end{figure}

\subsection{Sensitivity to Temperature under Interference}
\label{subsec:temperature}
Figure~\ref{fig:temp} shows how decoding temperature interacts with irrelevant audio during GSM8K reasoning. At low temperatures, both Qwen2.5-Omni-7B and Phi-4-Multimodal retain stable accuracy and relatively low influence rates, indicating that deterministic decoding reduces the impact of cross-modal interference. As temperature rises, however, accuracy begins to drop more steeply in the presence of silence, Gaussian noise, or FSD50K audio, revealing that stochastic sampling amplifies the destabilizing effect of irrelevant signals.

The influence rate highlights this compounding effect more clearly than accuracy alone. For both models, the influence rate is relatively low near greedy decoding but escalates sharply as temperature increases. This means that irrelevant audio reduces accuracy and actively makes predictions more volatile, flipping outcomes more frequently under higher sampling variance. The effect is particularly evident with real-world FSD50K inputs, where instability intensifies beyond what is observed with silence or synthetic noise.

Comparing models, Qwen2.5-Omni-7B sustains higher accuracy across all conditions and shows a slower rise in influence rate, indicating stronger resilience to interference. By contrast, Phi-4-Multimodal suffers sharper accuracy declines and a steeper increase in influence rate, suggesting greater susceptibility to stochasticity and irrelevant audio. These trends demonstrate that temperature is not a neutral hyperparameter under interference but a critical factor interacting with model design to shape robustness.

\section{Straightforward Mitigation Approaches}
\label{sec:mitigation}
\subsection{Methodology}
\label{subsec:method}
We evaluate two straightforward mitigation approaches to investigate whether simple strategies can alleviate the impact of irrelevant audio. The first approach is adding a mitigation prompt. Specifically, we prepend a short instructional phrase, \textit{``Focus on the text or audio that contains useful information.''} before each question. This prompt is designed to remind the model to pay explicit attention only to modalities that contribute to solving the task, thereby reducing the likelihood of being distracted by non-informative audio streams. The second approach is self-consistency. Instead of relying on a single decoding output, we generate 8 responses with sampling temperature 0.5 and aggregate the final answer by majority voting. This ensemble-style decoding reduces prediction volatility by smoothing out spurious shifts introduced by audio interference.

\input{Tables/mitigation}
\subsection{Results and Discussion}
\label{subsec:results}
Table \ref{tab:mitigation} compares the performance of the two mitigation strategies, prompting and self-consistency, under silence, Gaussian noise, and FSD50K interference. The influence rate is recalculated for each method relative to its own clean baseline.

The mitigation prompt yields only limited and inconsistent improvements. In some settings, it slightly lowers the influence rate, yet in others, it reduces accuracy or even amplifies instability. The variation across benchmarks and models shows that a single explicit instruction is insufficient to counteract cross-modal interference systematically.

In contrast, self-consistency consistently improves robustness relative to its baseline without mitigation. Accuracy increases across all interference types, and the influence rate decreases substantially. This approach stabilizes predictions and delivers more reliable outputs in the presence of irrelevant audio. The results demonstrate that aggregating multiple generations and selecting the majority answer effectively counteract instability introduced by audio interference.

In summary, mitigation prompts alone are insufficient to protect models against irrelevant audio, often yielding inconsistent or minor effects compared with their baseline. Self-consistency, on the other hand, reliably enhances both accuracy and robustness relative to its baseline, though at the cost of considerable computational overhead since multiple generations must be produced for each input.

\section{Conclusion}
\label{sec:conclusion}
Our study shows that irrelevant audio can interfere with how large audio-language models reason over text. Silence, noise, and environmental sounds disrupted performance, and the impact grew with longer duration, louder volume, and higher decoding temperatures. Even silence, often assumed neutral, proved disruptive, destabilizing outputs as much as synthetic noise. Larger models and specific architectures showed greater resilience, but none were fully robust. Mitigation experiments revealed that prompting was ineffective, whereas self-consistency improved stability but introduced substantial test-time compute overhead. These findings establish cross-modal interference as a key robustness challenge and call for more efficient fusion strategies to preserve reasoning quality in realistic multimodal settings.

\section{Acknowledgements}
We thank the reviewers for their insightful comments. This work was financially supported by the National Science and Technology Council (NSTC) in Taiwan, under Grant 114-2628-E-002-022. We thank to National Center for High-performance Computing (NCHC) of National Applied Research Laboratories (NARLabs) in Taiwan for providing computational and storage resources. This work was supported by the Ministry of Education (MOE) of Taiwan under the project Taiwan Centers of Excellence in Artificial Intelligence, through the NTU Artificial Intelligence Center of Research Excellence (NTU AI-CoRE). We are also grateful to Yu-Xiang Lin from National Taiwan University for his valuable advice and thoughtful discussions.

\bibliographystyle{IEEEbib}
\bibliography{strings,refs}

\end{document}

%% file: Tables/cross.tex
\begin{table}[hbtp]
\centering
\caption{Correctness change ratios across interference conditions. Results are reported on GSM8K, MMLU, and ARC-Challenge.}
\resizebox{\linewidth}{!}{
\begin{tabular}{l l c c c}
\toprule
\textbf{Model} & \textbf{Cond. Pair} & \textbf{GSM8K} & \textbf{MMLU} & \textbf{ARC} \\
\midrule
\multirow{3}{*}{Qwen2.5-Omni-3B}  
& silence/noise   & 0.057 & 0.078 & 0.048 \\
& silence/fsd50k  & 0.086 & 0.119 & 0.079 \\
& noise/fsd50k    & 0.084 & 0.120 & 0.077 \\
\midrule
\multirow{3}{*}{Phi-4-multimodal}
& silence/noise   & 0.083 & 0.159 & 0.113 \\
& silence/fsd50k  & 0.112 & 0.174 & 0.120 \\
& noise/fsd50k    & 0.095 & 0.164 & 0.114 \\
\bottomrule
\end{tabular}
}
\label{tab:cross}
\vspace{-10pt}
\end{table}

%% file: Tables/mitigation.tex
\begin{table}[t]
\centering
\caption{Comparing with several mitigation approach under different interference conditions.}
\resizebox{\linewidth}{!}{
\begin{tabular}{l l c c c c}
\toprule
\multirow{2}{*}{\textbf{Model}} & \multirow{2}{*}{\textbf{Condition}} & 
\multicolumn{2}{c}{\textbf{GSM8K}} & \multicolumn{2}{c}{\textbf{ARC-Challenge}} \\
\cmidrule(lr){3-4} \cmidrule(lr){5-6}
 & & Acc $\uparrow$ & IR $\downarrow$ & Acc $\uparrow$ & IR $\downarrow$ \\
\midrule
\multirow{4}{*}{Qwen2.5-Omni-3B} 
  & clean  & 0.7915 & - & 0.7782 & - \\
  & silence & 0.7915 & 0.1016 & 0.7765 & 0.0904 \\
  & noise   & 0.7862 & 0.0963 & 0.7713 & 0.1007 \\
  & fsd50k  & 0.7794 & 0.1001 & 0.7645 & 0.1143 \\
\midrule
\multirow{4}{*}{ + Prompt} 
  & clean  & 0.7779 & - & 0.7722 & - \\
  & silence & 0.7817 & 0.1054 & 0.7773 & 0.0802 \\
  & noise   & 0.7809 & 0.1168 & 0.7645 & 0.0930 \\
  & fsd50k  & 0.7817 & 0.1114 & 0.7696 & 0.1084 \\
\midrule
\multirow{4}{*}{ + Self-Consistency} 
  & clean  & 0.8552 & - & 0.8157 & - \\
  & silence & 0.8529 & 0.0432 & 0.8029 & 0.0555 \\
  & noise   & 0.8514 & 0.0478 & 0.7986 & 0.0597 \\
  & fsd50k  & 0.8628 & 0.0576 & 0.8080 & 0.0691 \\
\midrule
\multirow{4}{*}{Phi-4-multimodal} 
  & clean  & 0.8120 & - & 0.7884 & - \\
  & silence & 0.8021 & 0.1296 & 0.7628 & 0.1570 \\
  & noise   & 0.8029 & 0.1440 & 0.7440 & 0.1655 \\
  & fsd50k  & 0.7900 & 0.1463 & 0.7543 & 0.1621 \\
\midrule
\multirow{4}{*}{ + Prompt}
  & clean  & 0.8188 & - & 0.7816 & - \\
  & silence & 0.7900 & 0.1440 & 0.7619 & 0.1101 \\
  & noise   & 0.7892 & 0.1539 & 0.7543 & 0.1160 \\
  & fsd50k  & 0.7991 & 0.1365 & 0.7474 & 0.1263 \\
\midrule
\multirow{4}{*}{ + Self-Consistency} 
  & clean  & 0.8825 & - & 0.8370 & - \\
  & silence & 0.8688 & 0.0637 & 0.7961 & 0.1075 \\
  & noise   & 0.8688 & 0.0667 & 0.7739 & 0.1195 \\
  & fsd50k  & 0.8590 & 0.0720 & 0.7619 & 0.1280 \\
\bottomrule
\end{tabular}}
\label{tab:mitigation}
\vspace{-10pt}
\end{table}